\input harvmac.tex
\input psfig
\overfullrule=0pt

\def\simge{\mathrel{%
   \rlap{\raise 0.511ex \hbox{$>$}}{\lower 0.511ex \hbox{$\sim$}}}}
\def\simle{\mathrel{
   \rlap{\raise 0.511ex \hbox{$<$}}{\lower 0.511ex \hbox{$\sim$}}}}
 
\def\slashchar#1{\setbox0=\hbox{$#1$}           
   \dimen0=\wd0                                 
   \setbox1=\hbox{/} \dimen1=\wd1               
   \ifdim\dimen0>\dimen1                        
      \rlap{\hbox to \dimen0{\hfil/\hfil}}      
      #1                                        
   \else                                        
      \rlap{\hbox to \dimen1{\hfil$#1$\hfil}}   
      /                                         
   \fi}                                         %
\def\ts{\thinspace}

\def\ra{\rightarrow}

\def\ol{\bar}

\def\CC{{\cal C}}

\def\CO{{\cal O}}

\def\ecm{\sqrt{s}}
\def\shat{\hat s}

\def\atc{\alpha_{TC}}

\def\atro{\alpha_{\tro}}

\def\Ntc{N_{TC}}

\def\sutc{SU(\Ntc)}

\def\thw{\theta_W}
\def\kslash{\raise.15ex\hbox{/}\kern-.57em k}
\def\LTC{\Lambda_{TC}}
\def\LETC{\Lambda_{ETC}}

\def\tro{\rho_{T}}

\def\tropm{\rho_{T}^\pm}

\def\troz{\rho_{T}^0}
\def\tpi{\pi_T}
\def\tpipm{\pi_T^\pm}
\def\tpimp{\pi_T^\mp}
\def\tpip{\pi_T^+}
\def\tpim{\pi_T^-}
\def\tpiz{\pi_T^0}
\def\tpipr{\pi_T^{0 \prime}}

\def\toppi{\pi_t}

\def\condtbt{\langle \bar t t\rangle}
\def\condtct{\langle \bar T T\rangle}

\def\gev{{\rm GeV}}
\def\tev{{\rm TeV}}

\def\pb{{\rm pb}}

\def\fb{{\rm fb}}

\def\third{\textstyle{ { 1\over { 3 } }}}

\def\myfoot#1#2{{\baselineskip=14.4pt plus 0.3pt\footnote{#1}{#2}}}

\Title{\vbox{\baselineskip12pt\hbox{FERMILAB-PUB-96/075-T}
\hbox{BUHEP-96-9}
\hbox{hep-ph/9607213}}}
{Low-Scale Technicolor at the Tevatron}

\smallskip
\centerline{Estia Eichten\myfoot{$^{\dag }$}{eichten@fnal.gov}}
\smallskip\centerline{Fermi National Accelerator Laboratory}
\centerline{P.O.~Box 500 Batavia, IL 60510}
\centerline{and}
\smallskip
\centerline{Kenneth Lane\myfoot{$^{\ddag }$}{lane@buphyc.bu.edu}}
\smallskip\centerline{Department of Physics, Boston University}
\centerline{590 Commonwealth Avenue, Boston, MA 02215}
\vskip .3in

\centerline{\bf Abstract}

In multiscale models of walking technicolor, relatively light color-singlet
technipions are produced in $q \ol q$ annihilation in association with
longitudinal $W$ and $Z$ bosons and with each other. The technipions decay
as $\tpiz \ra b \ol b$ and $\tpip \ra c \ol b$. Their production rates are
resonantly enhanced by isovector technirho vector mesons with mass $M_W +
M_{\tpi} \simle M_{\tro} \simle 2 M_{\tpi}$. At the Tevatron, these
associated production rates are 1--10 picobarns for $M_{\tpi} \simeq
100\,\gev$. Such a low mass technipion requires topcolor-assisted
technicolor to suppress the decay $t \ra \tpip b$. Searches for $\tpi\tpi$
production will also be rewarding. Sizable rates are expected if $M_{\tro}
\simge 2M_{\tpi} + 10\,\gev$. The isoscalar $\omega_T$ is nearly degenerate 
with $\tro$ and is expected to be produced at roughly the same rate. The 
$\omega_T$ should have the distinctive decay modes $\omega_T \ra \gamma 
\tpiz$ and $Z \tpiz$.

\bigskip

\Date{7/96}

\vfil\eject

In the standard one-doublet Higgs model of electroweak symmetry breaking,
the cross section for production in $\ol p p$ collisions at $1.8\,\tev$ of
a $100\,\gev$ Higgs boson $H$ in association with $W^\pm$ bosons is about
$0.15\,\pb$. Even lower rates occur in nonminimal Higgs models, including
supersymmetric ones. Such small cross sections require luminosities of
1--$10\,\fb^{-1}$ to detect the signature $W^\pm + H \ra \ell^\pm + \ol b b
+ \slashchar{E}_T$
\ref\tevMM{``Report of the tev$\_$2000 Study Group on Future Electroweak
Physics at the Tevatron'', edited by D.~Amidei and R.~Brock,
D\O\ Note~2589 and CDF Note~3177, (1995).}.
In this note, we recall that very similar signatures occur in technicolor
models at rates that are 30 or more times greater than the Higgs rate,
large enough to be observed at the Tevatron now. In our 1989 paper
proposing multiscale technicolor
\ref\multia{K.~Lane and E.~Eichten, Phys. Lett. {\bf B222}, 274 (1989)},
we predicted that resonant production via isovector technirho ($\tro$) 
vector mesons of $W/Z + \tpi$, with $\tpi$ a
technipion of mass of $\sim 100\,\gev$ decaying to heavy quarks,
would occur at the Tevatron at the level of a few picobarns. Here, we
review our proposal and point out that such a low mass for the technipion
requires a new scenario such as topcolor-assisted technicolor
\ref\topcondref{Y.~Nambu, in {\it New Theories in Physics}, Proceedings of
the XI International Symposium on Elementary Particle Physics, Kazimierz,
Poland, 1988, edited by Z.~Adjuk, S.~Pokorski and A.~Trautmann (World
Scientific, Singapore, 1989); Enrico Fermi Institute Report EFI~89-08
(unpublished)\semi
V.~A.~Miransky, M.~Tanabashi and K.~Yamawaki, Phys.~Lett.~{\bf
221B}, 177 (1989); Mod.~Phys.~Lett.~{\bf A4}, 1043 (1989)\semi
W.~A.~Bardeen, C.~T.~Hill and M.~Lindner, Phys.~Rev.~{\bf D41},
1647 (1990).},
\ref\topcref{C.~T. Hill, Phys.~Lett.~{\bf 266B}, 419 (1991) \semi
S.~P.~Martin, Phys.~Rev.~{\bf D45}, 4283 (1992);
{\it ibid}~{\bf D46}, 2197 (1992); Nucl.~Phys.~{\bf B398}, 359 (1993);
M.~Lindner and D.~Ross, Nucl.~Phys.~{\bf  B370}, 30 (1992)\semi
R.~B\"{o}nisch, Phys.~Lett.~{\bf 268B}, 394 (1991)\semi
C.~T.~Hill, D.~Kennedy, T.~Onogi, H.~L.~Yu, Phys.~Rev.~{\bf D47}, 2940 
(1993).},
\ref\tctwohill{C.~T.~Hill, Phys.~Lett.~{\bf 345B}, 483 (1995).},
\ref\tctwoklee{K.~Lane and E.~Eichten, Phys.~Lett.~{\bf B352}, 382
(1995) \semi
K.~Lane, Boston University Preprint BUHEP--96--2, hep-ph/9602221, submitted
to Physical Review~D.}
to accommodate the top quark's large mass and prevent its decay to $\tpip
b$. If the custodial techni-isospin is approximately conserved, as we
expect, the isoscalar partner $\omega_T$ of the $\tro$ is nearly degenerate
with it and may be produced at a comparable rate. In addition, there may be
an isoscalar technipion, $\tpipr$, close in mass to to $\tpiz$. The main
decay modes of $\omega_T$ are expected to be $\gamma/Z + \tpiz$ and 
$\gamma/Z + \tpipr$, with $\tpiz$ and $\tpipr \ra \ol b b$, providing
spectactular signatures at the Tevatron.

Quark and lepton masses in technicolor are generated by broken extended
technicolor (ETC) gauge interactions
\ref\tcref{S.~Weinberg, Phys.~Rev.~{\bf D19}, 1277 (1979)\semi L.~Susskind,
Phys.~Rev.~{\bf D20}, 2619 (1979).},
\ref\etcDS{S.~Dimopoulos and L.~Susskind, Nucl.~Phys.~{\bf B155}, 237
(1979).},
\ref\etcEL{E.~Eichten and K.~Lane, Phys.~Lett.~{\bf 90B}, 125 (1980).}.
Because of the conflict between constraints on flavor-changing neutral
currents and the magnitude of ETC-generated masses, this classical version
of technicolor failed and was replaced a decade ago by ``walking''
technicolor
\ref\wtc{B.~Holdom, Phys.~Rev.~{\bf D24}, 1441 (1981);
Phys.~Lett.~{\bf 150B}, 301 (1985)\semi
T.~Appelquist, D.~Karabali and L.~C.~R. Wijewardhana,
Phys.~Rev.~Lett.~{\bf 57}, 957 (1986);
T.~Appelquist and L.~C.~R.~Wijewardhana, Phys.~Rev.~{\bf D36}, 568
(1987)\semi 
K.~Yamawaki, M.~Bando and K.~Matumoto, Phys.~Rev.~Lett.~{\bf 56}, 1335
(1986) \semi
T.~Akiba and T.~Yanagida, Phys.~Lett.~{\bf 169B}, 432 (1986).}.
In this kind of gauge theory, the strong technicolor coupling $\atc$ runs
very slowly for a large range of momenta, possibly all the way up to the
ETC scale, which must be several 100~TeV to suppress flavor-changing
effects. This slowly-running coupling permits quark and lepton masses as
large as a few~GeV to be generated from ETC interactions at this very high
scale.

Walking technicolor models require a large number of technifermions in
order that $\atc$ runs slowly. These fermions may belong to many copies of
the fundamental representation of the technicolor gauge group (here taken
to be $\sutc$), to a few higher dimensional representations, or to both.
This led us to argue in Ref.~\multia\ that both large and fundamental
representations participate in electroweak symmetry
breaking.\foot{Technicolor models with QCD-like dynamics cannot have a
large number of representations because they produce an $S$-parameter that
is too large and positive~\ref\pettests{A.~Longhitano, Phys.~Rev.~{\bf D22}
(1980) 1166; Nucl.~Phys.~{\bf B188}, (1981) 118\semi R.~Renken and
M.~Peskin, Nucl.~Phys.~{\bf B211} (1983) 93\semi
B.~W.~Lynn, M.~E.~Peskin and R~.G.~Stuart, in Trieste Electroweak 1985,
213\semi
M.~Golden and L.Randall, Nucl.~Phys.~{\bf B361} (1990) 3\semi B.~Holdom and
J.~Terning, Phys.~Lett.~{\bf B247} (1990) 88\semi
M.~E.~Peskin and T.~Takeuchi, Phys.~Rev.~Lett.~{\bf 65} (1990) 964\semi
A.~Dobado, D.~Espriu and M.~J.~Herrero, Phys.~Lett.~{\bf B255}
(1990) 405\semi
H.~Georgi, Nucl.~Phys.~{\bf B363} (1991) 301.}. Of course,
such models are already ruled out because they have flavor-changing neutral
currents that are too large~\etcEL, the problem that motivated walking
technicolor. The arguments in \pettests\ are based on scaling from QCD and on
chiral perturbation theory. They fail or are questionable in walking
technicolor models; see Ref.~\ref\ichep{K.~Lane, Proceedings of the 27th
International Conference on High Energy Physics, edited by P.~J.~Bussey and
I.~G.~Knowles, Vol.~II, p.~543, Glasgow, June 20--27, 1994.}.} The two
types of technifermion condense at widely separated scales
\ref\sepscales{S.~Raby, S.~Dimopoulos and L.~Susskind, Nucl.~Phys.~{\bf
B169}, 373 (1980).}.
The upper scale is set by the weak decay constant $F_\pi = 246\,\gev$.
Technihadrons associated with the lower scale may be so light, we said,
that they are within reach of Tevatron collider experiments.

Light-scale technihadrons generally consist of color singlets (discussed in
\multia) and nonsinglets (discussed in~\ref\multib{K.~Lane and
M.~V.~Ramana, Phys.~Rev.~{\bf D44}, 2678 (1991).}). In this note, we shall
be interested in the color singlets. We consider first the lightest
isotriplet of technirho vector mesons, $\tro^{\pm,0}$. We will discuss
their isoscalar counterpart, $\omega_T$, later. The $\tro$ decay into pairs
of isovector technipion states, $\Pi_T^{\pm,0}$. In general, the latter are
mixtures of the longitudinal weak bosons $W_L^\pm$, $Z_L^0$ and
mass-eigenstate (pseudo-Goldstone) technipions $\tpi^\pm, \tpiz$. In the
simplest parameterization, with just one light isotriplet of $\tpi$,
$\vert\Pi_T\rangle = \sin\chi \ts \vert
W_L\rangle + \cos\chi \ts \vert\tpi\rangle$, where $\sin\chi = F_T/F_\pi
\ll 1$ and $F_T$ is the decay constant of $\Pi_T$. The $\tro$ partial decay
rates are given by (assuming no other open decay channels)
\ref\dim{S.~Dimopoulos, S.~Raby and G.~Kane, Nucl.~Phys.~{\bf B182}, 77
(1981).},
\ref\ehlq{E.~Eichten, I.~Hinchliffe, K.~Lane and C.~Quigg,
Rev.~Mod.~Phys.~{\bf 56}, 579 (1984); Phys.~Rev.~{\bf D34}, 1547 (1986).}
\eqn\singwidth{
\Gamma(\tro \ra \pi_A \pi_B) = {2 \atro \CC^2_{AB}\over{3}} \ts
{\ts\ts p_{AB}^3\over {M^2_{\tro}}} \ts,}
where $p_{AB}$ is the technipion momentum and $\CC^2_{AB} = \sin^4\chi$,
$2\sin^2\chi \cos^2\chi$, $\cos^4\chi$ for $\pi_A \pi_B = W_L W_L$,
$W_L\tpi+\tpi W_L$, $\tpi\tpi$, respectively. For technifermions in the
fundamental representation of $SU(\Ntc)$, the $\tro \ra \tpi\tpi$
coupling $\atro$ obtained by {\it naive} scaling from QCD is given
by
\eqn\alpharho{\atro = 2.91 \left({3\over{\Ntc}}\right)\ts.}
In calculations, we take $\Ntc =4$.

Extended technicolor interactions couple technipions to quarks and leptons
with Higgs-like couplings. Technipions are then expected to decay into
heavy fermions:
\eqn\singdecay{\eqalign{
\tpiz &\ra \cases{b \ol b &if $M_{\tpi} < 2 m_t$,  \cr
t \ol t &if $M_{\tpi} > 2 m_t$; \cr} \cr
\tpip &\ra \cases{c \ol b \ts\ts\ts {\rm or} \ts\ts\ts c \ol s, \ts\ts
\tau^+ \nu_\tau 
&if $M_{\tpi} < m_t + m_b$, \cr
t \ol b &if $M_{\tpi} > m_t + m_b$. \cr} \cr}}

Now---and this is the important feature of walking technicolor---technipion
masses are enhanced by renormalizations that are so large that the $\tro
\ra \tpi\tpi$ channels may be closed or strongly suppressed. Thus, technirho
production at the Tevatron can lead to all of the processes
\eqn\singlet{\eqalign{
q \ol q' \ra W^\pm \ra \tropm &\ra \ts\ts W_L^\pm Z_L^0; \quad W_L^\pm
\tpiz, \ts\ts \tpipm Z_L^0; \quad \tpipm \tpiz \cr\cr
q \ol q \ra \gamma, Z^0 \ra \troz &\ra \ts\ts W_L^+ W_L^-; \quad W_L^\pm
\tpimp; \quad \tpip \tpim \cr}}
occurring at comparable rates despite the small angle~$\chi$.

To illustrate our points, we shall take $M_{\tpipm} \simeq M_{\tpiz} 
\simeq 110\,\gev$ and vary $M_{\rho^\pm_T}
\simeq M_{\rho^0_T}$ in our calculations. Note that, since $\tro$ has $I=1$,
these resonant processes do not lead to $Z^0\tpiz\ra \ell^+\ell^-/\ts \nu
\ol \nu + b \ol b$ final states. Such events must originate from
$\omega_T$, as we discuss later. Technirho events with a $Z^0$ and two
heavy quark jets have one $b$-jet and one $c$-jet.\foot{Note that $Z+\tpi$
events with a lost lepton can end up in the $W+\tpi$ sample, while
$W+\tpi$ with a lost lepton may be counted as $Z(\ra \nu\ol\nu) + \tpi$.}

The Drell-Yan processes in Eq.~\singlet\ have $\CO(\alpha^2)$ cross
sections and are unobservably small compared to backgrounds {\it unless}
the technirho resonances are not far above threshold, roughly $M_W +
M_{\tpi} \simle M_{\tro} \simle 2 M_{\tpi}$. This condition is favored by
multiscale technicolor. We estimate the $\tro\ra \pi_A \pi_B$
subprocess cross sections by vector meson dominance, taking the $\gamma, 
Z, W \ra \tro$ couplings from the condition that they reproduce $\gamma, 
Z, W \ra \pi_A \pi_B$ at zero energy. We obtain~\multia\
\eqn\singcross{
{d\hat\sigma(q_i \ol q_j \ra \tro^{\pm,0} \ra \pi_A\pi_B) \over{dz}} = 
{\pi \alpha^2 p_{AB}^3 \over{3 \shat^{5/2}}}
\ts {M^4_{\tro} \ts \ts (1-z^2) \over
{(\shat - M_{\tro}^2)^2 + \shat \Gamma_{\tro}^2}} \ts A_{ij}^{\pm,0}(\shat)
\CC^2_{AB} \ts,}
where $\shat$ is the subprocess energy, $z = \cos\theta$ is the $\pi_A$
production angle, and $\Gamma_{\tro}$ is the energy-dependent total width.
Ignoring Kobayashi-Maskawa mixing angles, the factors $A_{ij}^{\pm,0} =
\delta_{ij} A^{\pm,0}$ are
\eqn\afactors{\eqalign{
A^\pm &= {1 \over {4 \sin^4\thw}} \biggl({\shat \over {\shat -
M_W^2}}\biggr)^2
\ts, \cr
A^0   &= \biggl[Q_i + {2 \cos 2\thw \over {\sin^2 2\thw}} \ts
(T_{3i} - Q_i \sin^2\thw) \biggl({\shat \over {\shat - M_Z^2}}\biggr)
\biggl]^2
\cr 
&\ts + \biggl[Q_i - {2 Q_i \cos 2\thw \sin^2\thw \over{\sin^2
2\thw}} \ts \biggl({\shat \over {\shat - M_Z^2}}\biggr) \biggl]^2 \ts.}}
Here, $Q_i$ and $T_{3i}$ are the electric charge and third component of
weak isospin for quark~$q_{iL,R}$. In these processes, the $W+$dijet and
$Z+$dijet invariant masses exhibit a narrow peak not far above $200\,\gev$.
This peak will be smeared by energy resolutions, but it should be narrower
than expected from continuum production of $W/Z + H$.

Our final observation is that, after all this, ordinary multiscale
technicolor cannot accommodate the top quark. Its mass of $m_t \simeq
175\,\gev$
\ref\toprefs{F.~Abe, et al., The CDF Collaboration, Phys.~Rev.~Lett.~{\bf
73}, 225 (1994); Phys.~Rev.~{\bf D50}, 2966 (1994); Phys.~Rev.~Lett.~{\bf
74}, 2626 (1995) \semi
S.~Abachi, et al., The D\O\ Collaboration, Phys.~Rev.~Lett.~{\bf
74}, 2632 (1995).}
is much too large to be generated by ETC coupling of the top quark
to the low-scale technifermions~\multib. Furthermore, there cannot be a
charged technipion as light as $110\,\gev$ for, then, the top quark would
tend to decay into it. Taking the coupling of $\tpip$ to $\ol t_L b_R$ to
be $\sqrt{2} m^{ETC}_t/F_T$, where $m^{ETC}_t$ is the ETC contribution to 
the top-quark mass, the top decay rate to $\tpip b$ is
\eqn\tpibrate{\Gamma(t \ra \tpip b) =
{(m_t^2 - M_{\tpi}^2)^2 \over {16 \pi m_t F_T^2}}\ts
\left( {m^{ETC}_t \over {m_t}}\right)^2 \ts.}
For $m^{ETC}_t \simeq m_t$ and $F_T = 40\,\gev$, a typical value in
multiscale models, $\Gamma(t \ra \tpip b) \simeq 25\,\gev \simeq 15
\Gamma(t \ra W^+ b)$; this is ruled out
\ref\twbrate{J.~Incandela, Proceedings of the 10th Topical Workshop on
Proton-Antiproton Collider Physics, Fermilab, edited R.~Raja and J.~Yoh,
p.~256 (1995).}.
Topcolor-assisted technicolor (TC2) resolves these problems.

In TC2~\tctwohill, as in top-condensate models of electroweak symmetry
breaking~\topcondref,\topcref, the large top quark mass is generated by
strong ``topcolor'' gauge interactions. Thus, there can be low-scale
technifermions and the ETC scale can be $\CO(100\,\tev)$ for all fermions.
Then, $m^{ETC}_t$ is only a few~GeV, so that the branching fraction $B(t
\ra \tpip b)$ is small.\foot{Top-pions $\toppi$ arising from top-quark
condensation do couple to $m_t$. Their mass arises from the ETC
contribution: $M^2_{\pi_t} \simeq m^{ETC}_t \condtbt/F^2_t$, where $F_t
\simeq 70\,\gev$~\tctwohill. This is sufficient to make $M_{\toppi} \simge 
m_t$.  Mixing of top-pions with technipions is expected to be small, 
of order $\condtbt/\condtct_{ETC} \simeq \LTC/\LETC \simle 10^{-2}$
\ref\balaji{B.~Balaji, ``Top-Pion--Technipion Mixing in Natural 
Topcolor-Assisted Technicolor'', Boston University preprint in preparation.}.}

Preliminary models of topcolor-assisted technicolor were developed in
Ref.~\tctwoklee. They differ from multiscale technicolor models in that
they do not contain technifermions in higher representations and the
associated widely separated scales. However, there are many copies of the
fundamental representation (some of which also transform under color or
topcolor $SU(3)$). Thus, the net effect is the same; ignoring $SU(3)$
effects, $F_T \simeq F_\pi/\sqrt{N_D}$, where $N_D$ is the number of
technifermion weak isodoublets. In the model of the second paper in
\tctwoklee, $N_D = 9$, so that $F_T = 82\,\gev$ (or somewhat less because
of color effects) for the lightest color-singlet technipions.${^1}$

We have used Eq.~\singcross\ with $\sin\chi = \third$ to compute the $\tro$
production cross sections for $M_{\tpi} = 110\,\gev$ and $M_{\tro} =
195$--$250\,\gev$. The individual decay channel cross sections are shown
for $p \ol p$ collisions at $\ecm = 1.8\,\tev$ in Fig.~1 (multiplied by a
$K$-factor of 1.5, appropriate for Drell-Yan processes at the Tevatron). No
cuts were put on the technipion directions. These plots illustrate several
features we expect to be general:

\item{$\circ$} Except near $W\pi_T$ threshold, the increase in $WZ$ and $WW$
production is small.

\item{$\circ$} The inclusive $W\tpi$ rate is 5--$10\,\pb$ and the $Z\tpi$
rate is 1--$3\,\pb$ for $M_{\tpi} + M_W \simle M_{\tro} \simle 2M_{\tpi}$.
The ratio $\sigma(W_L\tpi) / \sigma(Z_L\tpi) = 2$--3 is about the same as
expected for $\sigma(W H) / \sigma(Z H)$. Because of threshold singularities,
these rates can be changed significantly by modest isospin splittings in the
technihadron masses. The cross sections increase by 15--20\% when the Tevatron
energy is increased to 2~TeV.

\item{$\circ$} Because the technirho is narrow, the total $p p^\pm$ cross
section at $\ecm$ into the channel $\pi_A\pi_B$ is 
\eqn\sigtot{\eqalign{
&\sigma(p p^\pm \ra \tro \ra \pi_A \pi_B) \simeq {2 \pi^2 \over {3s}} \ts
\sum_{i,j} \ts {\Gamma(\tro \ra q_i\ol q_j) \ts \Gamma(\tro \ra \pi_A\pi_B)
\over {M_{\tro} \ts \Gamma_{\tro}}} \cr
&\qquad \times \int d \eta_B \ts
\left\{f_{q_i}^p\biggl({M_{\tro}\over{\sqrt{s}}} e^{\eta_B}\biggr)
\ts f_{\ol q_j}^{p^\pm}\biggl({M_{\tro} \over{\sqrt{s}}} e^{-\eta_B}\biggr) +
\ts\ts(q_i \leftrightarrow \ol q_j) \right\}\ts, \cr}}
where $f_{q_i}^{p^\pm}$ is the $q_i$ distribution function in $p^\pm$ at 
$Q^2 = M^2_{\tro}$. At the Tevatron, these distributions favor $\tro^\pm$
production over $\tro^0$ by a factor of 2--3 over the range of $M_{\tro}$
considered.

\item{$\circ$} For $M_{\tro} \ge 2 M_{\tpi} + 10\,\gev$, the dominant
process is $\tpipm\tpiz$ production. The crossover point depends to some
extent on the suppression factor $\tan\chi$, but we don't expect it to be
much different from this. A search for the $\tpipm\tpiz$ channel will be
rewarding.

Finally, we turn to the $\omega_T$. The walking technicolor enhancement of
technipion masses almost certainly closes off the isospin-conserving decay
$\omega_T \ra \Pi^+_T \Pi^-_T \Pi^0_T$. Even the triply-suppressed mode
$W^+_L W^-_L Z_L$ has little or no phase space for the $M_{\omega_T}$-range
we are considering. Thus, we expect the decays $\omega_T \ra \gamma
\Pi_T^0$, $Z \Pi_T^0$, and $\Pi_T^+ \Pi_T^-$. When written in terms of 
mass eigenstates, these modes are $\omega_T \ra \gamma \tpiz$, $\gamma
Z_L$, $Z \tpiz$, $Z Z_L$; $\gamma \tpipr$, $Z \tpipr$; and $W^+_L W^-_L$,
$\tpipm W^\mp_L$, $\tpip\tpim$.\foot{The modes $\omega_T \ra \gamma Z_L$,
$Z Z_L$ were considered for a one-doublet technicolor model in
Ref.~\ref\cg{R.~S.~Chivukula and M.~Golden, Phys.~Rev.~{\bf D41}, 2795
(1990).}. We have estimated the rates for the isospin-violating decays
$\tro \ra \gamma \tpiz$, $Z \tpiz$ and find them to be negligible unless
the mixing angle $\chi$ is very small.} It is not possible to estimate the
relative magnitudes of the decay amplitudes without an explicit model of
the $\omega_T$'s constituent technifermions. Judging from the decays of the
ordinary $\omega$, we expect $\omega_T \ra \gamma \tpiz (\tpipr)$, $Z \tpiz
(\tpipr)$ to be dominant, with the former mode favored by phase space.

The $\omega_T$ is produced in hadron collisions just as the $\tro^0$, via
its vector-meson-dominance coupling to $\gamma$ and $Z^0$. For
$M_{\omega_T} \simeq M_{\tro}$, the $\omega_T$ production cross section
should be approximately $|Q_U + Q_D|^2$ times the $\tro^0$ rate, where
$Q_{U,D}$ are the electric charges of the $\omega_T$'s constituent
technifermions. The principal signatures for $\omega_T$ production, then,
are $\gamma + \ol b b$ and $\ell^+\ell^-$ (or $\nu \ol \nu$) $+ b \ol b$,
with $M_{\ol b b} = M_{\tpi}$.

The model of color-singlet technihadron production we have 
discussed here is an oversimplification, but one that captures some of the
essence of modern models of technicolor.
As we did in Ref.~\multia, we urge a search for $W/Z + \tpi \ra$ isolated
high-$p_T$ leptons $+$ heavy quark jets. If such events are found, the
$W/Z+jj$ mass spectrum should exhibit a narrow $\tro$ peak consistent with
resolution. Sidebands---$M_{Wjj}$ with $M_{jj}$ outside the ``$M_{\tpi}$''
bins---should not exhibit the peak unless it turns out to be kinematic in
nature. Because of the quadratic ambiguity in reconstructing the $W$ in its
$\ell\nu$ decay, it seems best to plot the low-mass solution versus the
high-mass one. We also urge a search for $\tpi\tpi$ production. This may
not be possible now, but a search with the high-luminosity data of Tevatron
Run~II should be conclusive. If a $\tpi$ candidate is found, it will be
important to determine whether $c$-quarks as well as $b$-quarks occur in
its decay. Higher luminosity and more robust heavy-flavor tagging can make
this possible. Finally, there should be an isoscalar $\omega_T$ nearly 
degenerate with $\tro^0$, generically produced with a comparable cross 
section, and with spectacular $\gamma \ol b b$ and $Z \ol b b$ decay 
signatures. If these technihadrons are found, they will be just the first 
of a very large family.

\bigskip

EE's research is supported by the Fermi National Accelerator Laboratory,
which is operated by Universities Research Association, Inc., under
Contract~No.~DE--AC02--76CHO3000. KL's research is supported in part by the
Department of Energy under Grant~No.~DE--FG02--91ER40676. We gratefully
acknowledge the hospitality of the Aspen Center for Physics where this 
work was completed.

\listrefs


\vfil\eject
\includegraphics{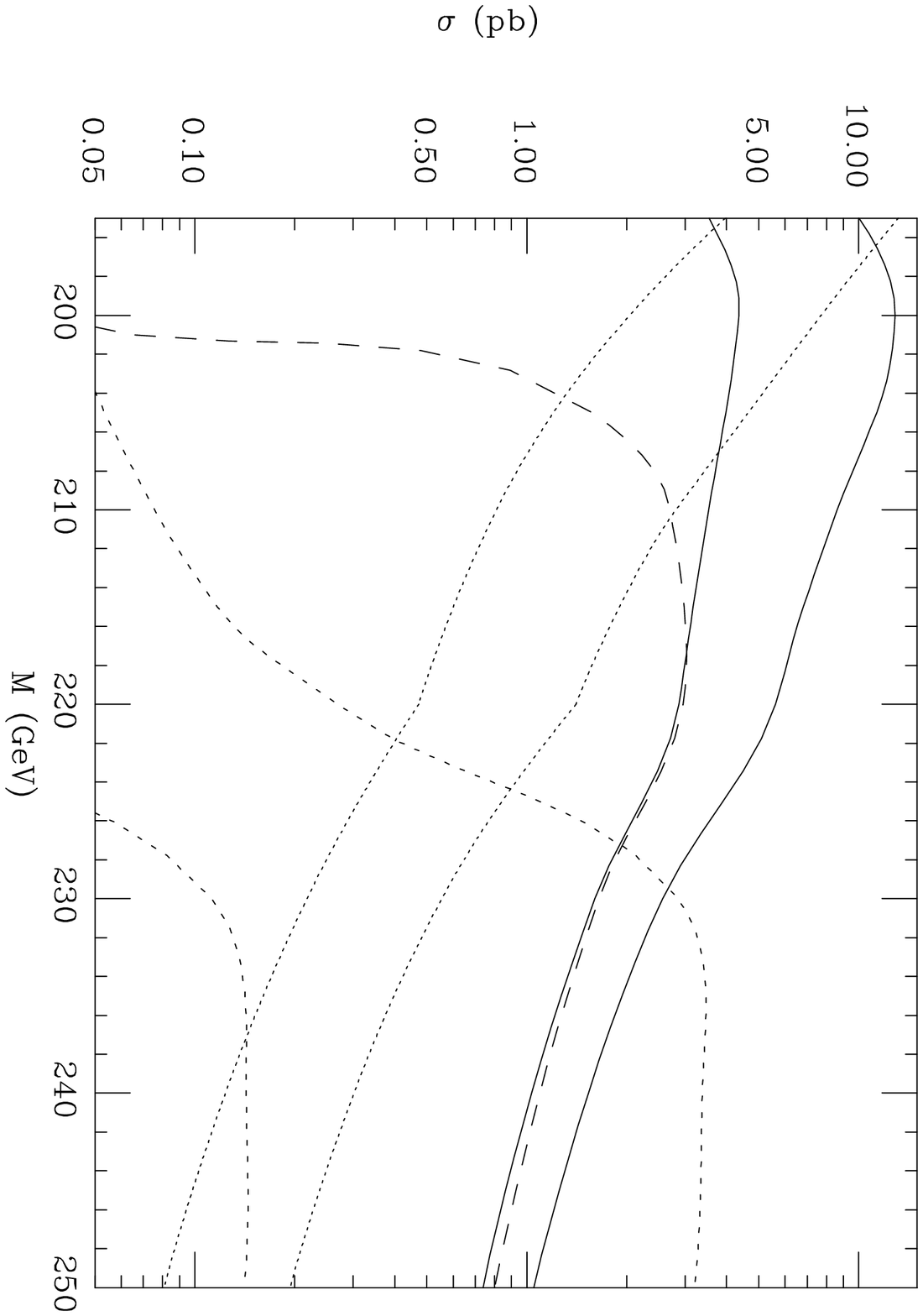}

\voffset5.5in

\noindent Figure 1. 
Total $WW$, $W\tpi$ and $\tpi\tpi$ cross sections in $\ol p p$
collisions at $1.8\,\tev$, as a function of $M_{\tro}$ for $M{\tpi} =
110\,\gev$. The model described above Eq.~\singwidth\ is used with $\sin\chi
= \third$. The curves are $W^\pm Z^0$ (upper dotted) and $W^+W^-$ (lower
dotted); $W^\pm \tpiz$ (upper solid), $W^\pm \tpi^\mp$ (lower solid), and
$Z^0\tpi^\pm$ (long dashed); $\tpi^\pm \tpiz$ (upper short dashed) and
$\tpip\tpim$ (lower short dashed). EHLQ set 1 distribution functions~\ehlq\
were used and cross sections were multiplied by a
$K$-factor of 1.5, as appropriate for Drell-Yan processes.

\bye